\documentclass[journal]{IEEEtran}
\usepackage{cite}

\ifCLASSINFOpdf
\else
\fi
\usepackage{amsmath}
\usepackage{amssymb}
\usepackage{bm} 

\newcommand{\norm}[1]{\left\lVert#1\right\rVert}

\usepackage{siunitx}						
\DeclareSIUnit\byte{B}

\usepackage{csquotes}
\usepackage{cleveref}

\usepackage{booktabs}
\usepackage{tabularx}
\usepackage{multirow}

\usepackage{graphicx}

\usepackage{pgfplots}
\pgfplotsset{compat=1.18}
\usepackage{pgfplotstable}
\usetikzlibrary{plotmarks}

\definecolor{aaubluedark}{RGB}{33,26,82}
\definecolor{aaubluelight}{RGB}{89,79,191}
\definecolor{aaugrey}{RGB}{84,97,110}
\definecolor{aausup1}{RGB}{106,119,132}
\definecolor{aausup2}{RGB}{162,172,182}
\definecolor{aausup3}{RGB}{222,223,226}
\definecolor{aausec1}{RGB}{157,187,29}
\definecolor{aausec2}{RGB}{94,150,149}
\definecolor{aausec3}{RGB}{223,103,82}

\usepackage{array}
\usepackage{url}

\begin{document}
%
\title{Noise-Robust Target-Speaker Voice Activity Detection Through Self-Supervised Pretraining}
%
%
%

\author{Holger Severin Bovbjerg, Jan Østergaard, Jesper Jensen, Zheng-Hua Tan
\thanks{H. S. Bovbjerg and J. Østergaard is with the Department of Electronic Systems, Aalborg University, Aalborg 9220, Denmark,
e-mail: hsbo@es.aau.dk; jo@es.aau.dk.}
\thanks{J. Jensen is with the Department of Electronic Systems, Aalborg University, Aalborg 9220, Denmark, and also with Oticon A/S, Smørum 2765, Denmark (e-mail: jje@es.aau.dk).}
\thanks{Z.H. Tan is with Department of Electronic Systems, Aalborg University, Aalborg 9220, Denmark, and also with Pioneer Centre for AI, Denmark.}
\thanks{This work has been submitted to the IEEE/ACM Transactions on Audio, Speech, and Language Processing for possible publication. Copyright may be transferred without notice, after which this version may no longer be accessible.}
}

\maketitle

\begin{abstract}
Target-Speaker Voice Activity Detection (TS-VAD) is the task of detecting the presence of speech from a known target-speaker in an audio frame.
Recently, deep neural network-based models have shown good performance in this task.
However, training these models requires extensive labelled data, which is costly and time-consuming to obtain, particularly if generalization to unseen environments is crucial. 
To mitigate this, we propose a causal, Self-Supervised Learning (SSL) pretraining framework, called Denoising Autoregressive Predictive Coding (DN-APC), to enhance TS-VAD performance in noisy conditions. 
We also explore various speaker conditioning methods and evaluate their performance under different noisy conditions. 
Our experiments show that DN-APC improves performance in noisy conditions, with a general improvement of approx. \SI{2}{\percent} in both seen and unseen noise. 
Additionally, we find that FiLM conditioning provides the best overall performance.
Representation analysis via tSNE plots reveals robust initial representations of speech and non-speech from pretraining. 
This underscores the effectiveness of SSL pretraining in improving the robustness and performance of TS-VAD models in noisy environments.
\end{abstract}


%
\IEEEpeerreviewmaketitle

\section{Introduction}
%
%
%
%
\IEEEPARstart{D}{etecting} the presence of speech is one of the most fundamental human skills. 
For decades, people have been developing Voice Activity Detection (VAD) methods to enable computers to do the same.
These methods include both unsupervised approaches \cite{TAN_rVAD} and supervised data-driven approaches \cite{dinkel2021voice, li2024robust}. VAD methods serve many purposes, such as noise reduction, data compression and speech analysis. 
Their applications include automatic speech recognition (ASR), hearing aids, voice assistants, surveillance \cite{chettri2020voice} and multimedia indexing \cite{Vryzas_VAD_indexing}.
Recent advances in signal processing, driven by the advent of data-driven approaches using deep neural networks (DNNs), have made it possible to build speech models which can focus on a particular target speaker \cite{Zmolikova_neural_ts_extraction}.
These models use various prior information about the target speaker, such as speaker characteristics \cite{zmolikova_speakerbeam, ding20_PVAD1, ding22_PVAD2, Peng_ts_extraction_using_ssl, Medennikov_TS_VAD, Maokui_TS_VAD_improved_i_vector, dongmei_tsvad_transformer, dongmei_profile_error_tolerant_tsvad}, visual information \cite{Sato_MM_att_fusion_ts_extraction, Ochiai_MM_speaker_beam, Lin_AV_sepformer, Pan_universal_SE_visual} or brain signals \cite{Ceolini_brain_informed_speech_separation, mirkovic_ts_extraction_concealed_eeg, teoh_eeg_decoding_of_ts, simon_cortical_auditory_attention_decoding} to provide a \textit{personalized} output.
Such models can be used in applications like enabling hearing aids to focus on a specific target speaker, and filtering the voice input for an ASR system such that only the target-speaker's words are decoded. 

In this work, we focus on Target-Speaker VAD (TS-VAD) models which use speaker characteristics information to \textit{personalize} the model output, enabling the detection of the presence of target-speech in an audio frame.
Such systems commonly use a target-speaker embedding vector computed from prerecorded enrolment speech by the target-speaker.
Noteworthy examples of such models are PersonalVAD (PVAD) \cite{ding20_PVAD1, ding22_PVAD2}, and TS-VAD \cite{Medennikov_TS_VAD, dongmei_tsvad_transformer, dongmei_profile_error_tolerant_tsvad} which have been used for speaker diarization. 
However, training these models requires a substantial amount of diverse training data, especially if generalization to unseen noisy conditions is crucial. 
While unlabelled audio data is generally easy to obtain, training a target-speaker VAD model requires suitable data labels.
These labels must both contain information about the time-stamps for speech activity and the  identity of the speaker at all times.
Obtaining such labels is a costly and time-consuming process, as it usually involves human supervision.
As a result, the cost of obtaining labelled data for training is currently a performance bottleneck, limiting the performance as well as wide-spread adoption of target-speaker VAD models.

In an effort to mitigate the need for labelled data to train DNNs, the paradigm of Self-Supervised Learning (SSL) has gained popularity \cite{Abdelrahman_ssl_speech_review}. 
In SSL, the DNN is trained to solve a self-supervised \textit{pretext} task on a large pool of unlabelled data.
Contrary to supervised learning, self-supervised pretext tasks derive the supervision label from the raw data itself.
For example, one might mask some part of an audio sequence and train a model to reconstruct the masked part, using the masked part as the ground truth for supervision. 
In the process of learning to predict the hidden part, the model learns to extract information from the input. 
Hopefully, the information used for solving the pretext task is also useful for solving the downstream task.
After pretraining on the pretext task using the unlabelled data, the model is \textit{fine-tuned} for the \textit{downstream} task in a standard supervised fashion, using the small amount of labelled data available.

Large pretrained SSL models have shown a remarkable ability to generalize to various speech tasks and currently form the foundation for many state-of-the-art speech models \cite{baevski_wav2vec2, HuBERT_paper, Chen_WavLMLS}.
The large pretrained SSL models are used as a general purpose audio feature extractor, which can be fine-tuned for a broad range of down stream tasks, e.g., ASR, speaker recognition or speech synthesis. 
However, the area of SSL for pretraining of low-complexity task-specific models is currently underexplored.

In our recent work, we proposed using a self-supervised pretraining stage to improve the performance of a sub-100k parameter Long Short-Term Memory (LSTM)-based TS-VAD model in noisy conditions \cite{bovbjerg_ssl_pvad}.
Our goal was not to develop a general audio representation model, but to benefit from SSL to create a model which is more robust to adverse conditions, such as background noise.
Our experiments showed that SSL pretraining not only yielded an improvement in clean conditions, but also significantly improved the performance of the model in noisy conditions.
This work extends on our previous work as follows:

\begin{itemize}
    \item We present a thorough analysis of the TS-VAD problem, providing deeper insights into the various aspects to consider when designing a TS-VAD model.
    \item Our previous work used a separate VAD and speaker verification model, and only focused on pretraining of the VAD for improving noise robustness. 
    In this work, we show that pretraining improves both the models' ability to distinguish speech from non-speech and its ability to distinguish target-speech from non-target speech. Additionally, we include experiments using various speaker-conditioning methods for joint modelling of voice activity and target-speaker presence.
    \item We use an updated causal Conformer-based architecture, which was found to improve performance \cite{ding22_PVAD2}, instead of the previous LSTM based model.
    \item We explore the learned representations of both supervised and self-supervised TS-VAD models, and show that the pretrained model is able to distinguish noisy speech from noise even without fine-tuning. Additionally, we show that the fine-tuned model has better class separation in noise compared to the supervised baseline. 
\end{itemize}


The rest of the paper is structured as follows:
First, the TS-VAD problem is introduced formally in \Cref{sec:ts-vad}, along with an analysis of different processing approaches.
In \Cref{sec:speaker_conditioning}, we present an analysis of various speaker-conditioning methods, exploring different methods to incorporate target-speaker information into the model's prediction.
In \Cref{sec:ssl}, we give a brief introduction to SSL and present our DN-APC SSL framework, which we use for pretraining of our TS-VAD models.
\Cref{sec:experiments} provides the experimental settings, including data sets and model setup. In \Cref{sec:results}, we present the experimental results as a series of \enquote{sub studies}. 
First, we compare our supervised baseline TS-VAD model with TS-VAD models pretrained using DN-APC.
Here, we show that the SSL pretrained models not only perform better in clean conditions, but are also more robust in noisy conditions than our supervised baseline.
Secondly, we study how different speaker-conditioning methods affect the TS-VAD performance.
Lastly, we carry out an explanation study of the learned representations, in which we try to quantify what makes the pretrained models more robust to noise than the purely supervised models.
We conclude on our work and present key takeaways from our study in \Cref{sec:conclusion}.




\section{Target-speaker Voice Activity Detection}\label{sec:ts-vad}
The problem of detecting the voice activity of a target-speaker based on an observed audio signal given some prior information about the target-speaker has been studied in multiple works.
In this work, we refer to the general problem as target-speaker voice activity detection (TS-VAD), although it has been referred to as both TS-VAD \cite{Medennikov_TS_VAD, dongmei_tsvad_transformer, Maokui_TS_VAD_improved_i_vector} and personalized voice activity detection (PVAD) \cite{ding20_PVAD1, ding22_PVAD2, bovbjerg_ssl_pvad} in the literature.

\subsection{Signal model}
We assume a discrete-time observed signal $x[t]$, with $t$ denoting the sample at time $\frac{t}{F_\mathrm{s}}$ and $F_\mathrm{s}$ being the sampling frequency for the observed signal.
The observed signal $x[t]$ might contain other sources than the target-speaker, such as background noise and other speakers, thus the signal can be modelled as
\begin{equation}
    x[t] = s^{\mathrm{t}}[t] + s^{\mathrm{nt}}[t] + v[t] \label{eq:signal_model}
\end{equation}
where $s^{\mathrm{t}}[t]$ represents the clean target-speaker signal, $s^{\mathrm{nt}}[t]$ represents the clean non-target-speaker sources and $v[t]$ represents sources of additive background noise.

The analysis of the observed signal is normally done on a per-frame basis, using an analysis window $w[t]$ of length $L$ which is shifted over the observed signal $x[t]$ in steps of $M$ samples, yielding observed frames
\begin{equation}
    x_n[t] = x[t + nM] \cdot w[t],\quad t=0,1,\dots,L-1, n\in \mathbb{Z},
\end{equation}
where ${x}_n[t]$ represents the $t$'th sample of the $n$'th frame from the observed signal.

The frames are usually passed through a feature extractor \cite{graf_vad_features_analysis}, e.g., a Discrete Fourier Transform (DFT) followed by a Mel-scale filterbank as in \cite{ding20_PVAD1}, to extract features 
\begin{equation}
    \bm{y}_n = f_\mathrm{FE}(\bm{x}_{n}),
\end{equation}
where $\bm{y}_n \in \mathbb{R}^{ D_\mathrm{feat} \times 1}$ denotes the extracted feature vector for the $n$'th frame, $f_\mathrm{FE}$ represents some feature extraction function, and $\bm{x}_n \in \mathbb{R}^{L \times 1} $ denotes the vector consisting of all $L$ elements of the $n$'th frame. 
In some cases, the features for frame $n$ are computed from multiple frames such that $\bm{y}_n = f_\mathrm{FE}(\bm{X}_{n})$, with
$\bm{X}_n = [\bm{x}_{n-c_\mathrm{left}},\dots, \bm{x}_{n}, \dots, \bm{x}_{n+c_\mathrm{right}}]$ given left context $c_\mathrm{left}$ and right context $c_\mathrm{right}$.
The feature extraction function $f_\mathrm{FE}$ can be a fixed function, such as a Mel feature extractor, or a function with learnable parameters, e.g., a Convolutional Neural Network (CNN) feature extraction network.

\subsection{Target-speaker information}
To differentiate the target-speaker from non-target-speakers, some information about the desired target-speaker is needed.
For instance, this could be information about the voice characteristics of the target-speaker \cite{ding20_PVAD1}, visual queues of the target \cite{Ceolini_brain_informed_speech_separation}, or information decoded from brain signals \cite{Ceolini_brain_informed_speech_separation}. 
However, in this work, we constrain ourselves to the case where using voice characteristics computed from \textit{enrolment} speech from the target-speaker. 

The voice characteristics of the target-speaker is usually obtained by passing some \textit{enrolment} speech from the target-speaker through an embedding model $f_\mathrm{emb}$. 
Until recently, the most popular method for implementing the embedding model $f_\mathrm{emb}$ was \textit{i-vector} \cite{gupta_ivector}. 
The \textit{i-vector} approach is based on statistics from a Gaussian mixture model trained on speech data from multiple speakers.
In recent years, DNN-based speaker embedding models, such as \textit{d-vector} \cite{variani_dvector, Wan_GE2E} and \textit{x-vector} \cite{snyder_xvector}, have proven to outperform the classical \textit{i-vector} approach in most scenarios. 
The \textit{d-vector} and \textit{x-vector} approaches both train a DNN to discriminate different speakers in the training data, and the resulting model is then used to extract speaker characteristics.
Similar for all methods, is that they try to summarize the target-speaker characteristics into an embedding vector $\bm{e}^{\mathrm{t}} \in \mathbb{R}^{1 \times D_\mathrm{emb}}$ such that
\begin{equation}
    \bm{e}^{\mathrm{t}} = f_\mathrm{emb}(s^\mathrm{t}_\mathrm{enrol}),
\end{equation}
where $\bm{s}^\mathrm{t}_\mathrm{enrol}$ represents one or more sequences of recorded enrolment speech from the target-speaker.

Thereby, the problem can be stated as predicting the voice activity related to the target-speaker signal $s^\mathrm{t}[t]$ of the observed frame $\bm{y}_n$, conditioned on the target-speaker embedding vector $\bm{e}^{\mathrm{t}}$ such that predicted voice activity becomes
\begin{equation}
    z^{\mathrm{ts}}_n = f_\mathbf{\theta}(\bm{y}_n;\bm{e}^{\mathrm{t}}) \label{eq:target_speech_prediction}
\end{equation}
where $z^{\mathrm{ts}}_n$ denotes the probability of target-speaker voice activity in the $n$'th frame and $f$ denotes some function parameterized by parameters $\mathbf{\theta}$.

Ideally, the function $f$ should have the following two properties:
\begin{enumerate}
    \item It should be able to discriminate between speech and non-speech.
    \item It should be able to determine whether a frame of speech was uttered by the target-speaker, given the target information available in $\bm{e}^{\mathrm{t}}$.
\end{enumerate}

While (1) can be solved from $\bm{y}_n$ directly using traditional VAD algorithms, (2) relies on both using $\bm{y}_n$ and $\bm{e}^{\mathrm{t}}$ to resolve whether $\bm{y}_n$ contains speech with similar characteristics as those embedded in $\bm{e}^{\mathrm{t}}$.
Therefore, the problem can be divided into two tasks, namely VAD and Speaker Verification (SV). 
Estimating $f$ can thus be posed as solving these two subtasks, which might be done in parallel, sequentially, or jointly.

\subsection{Parallel Processing}
A naive approach to the TS-VAD problem, is to use separate VAD and SV models, run the speech signal through both in parallel, and combine the scores.
In\,\cite{ding20_PVAD1} the authors denote the parallel approach as \textit{score combination}. 
Here, the input is passed through two models. 

One model, $f_\mathrm{VAD}$, models speech activity,
\begin{equation}
    [z^\mathrm{ns}_n, z^\mathrm{s}_n] = f_\mathrm{VAD}(\bm{y}_n),
\end{equation}
where $z^\mathrm{ns}_n$ and $z^\mathrm{s}_n$ represent the predicted probabilities for no speech and speech, respectively.

The second model, $f_\mathrm{SV}$, models speaker similarity.
This model is typically implemented using the same model used to generate $\bm{e}^{\mathrm{t}}$, i.e., $f_\mathrm{emb}$.
Specifically, a speaker embedding of the current frame is generated and its cosine similarity with the target speaker embedding is computed,
\begin{equation}
    \mathrm{sim}^\mathrm{t} = f_\mathrm{SV}(\bm{y}_n, \bm{e}^{\mathrm{t}}) = \cos(\bm{e}_n, \bm{e}^{\mathrm{t}})
\end{equation}
where $\bm{e}_n=f_\mathrm{emb}(\bm{y}_n)$ is the speaker embedding of the $n$'th frame.

Using $z^\mathrm{s}_n$ and $\mathrm{sim}^\mathrm{t}$, the prediction of $z^{\mathrm{ts}}_n$ can be solved as a binary classification problem.
However, in\,\cite{ding20_PVAD1}, it was empirically found that posing the problem as a threefold classification problem, distinguishing between non-speech (ns), target-speaker speech (ts) or non-target-speaker speech (nts) lead to improved performance.
In this case, each frame is classified by combining the scores such that
 \begin{equation}
    z_n^k =
    \begin{cases}
        z^\mathrm{ns}_n, & k = \mathrm{ns}, \\ 
        \mathrm{sim}^\mathrm{t} z^\mathrm{s}_n, & k = \mathrm{ts}, \\ 
        (1 - \mathrm{sim}^\mathrm{t})z^\mathrm{s}_n, & k = \mathrm{nts},
    \end{cases} \label{eq:score_combination}
\end{equation}
with $z^k_n$ denoting the output corresponding to class $k$ at the $n$'th frame.

\subsection{Cascaded Processing}
Using a VAD and an SV model in parallel can be computationally expensive, especially due to the relatively large size of SV models compared to VAD models. 
A cascade processing approach can mitigate this problem by only running the SV model when speech is present.
In this setup, the VAD model is first used to determine whether an input frame contains speech. 
If so, the speaker verification system is utilized to determine whether that frame of speech originates from the target-speaker.

While the cascaded processing approach can mitigate some of the computational cost of running the SV model during runtime, it still requires explicitly solving each sub problem using two individual models, which might potentially carry out redundant processing. 
Additionally, it adds to the algorithmic delay, as the speaker similarity scores are not computed until the VAD score is available.

\subsection{Joint processing}
The joint processing approach completely removes the need to run a dedicated SV system during runtime.
This is done by combining $f_\mathrm{VAD}$ and $f_\mathrm{SV}$ into a single model. 
In this setup, no intermediate scores are computed and instead the model handles both tasks jointly, i.e., the VAD and SV scores are not explicitly computed by themselves and combined using \labelcref{eq:score_combination}.
The joint model can be expressed as
\begin{equation}
    [z_n^\mathrm{ns}, z_n^\mathrm{ts}, z_n^\mathrm{nts}]=f_\mathrm{TSVAD}(\bm{y}_n; \bm{e}^{\mathrm{t}})
\end{equation}
where $f_\mathrm{TSVAD}$ is a function that jointly models speech activity and speaker similarity.

Joint processing can be enabled through the use of speaker-conditioning, which is used to condition a model's prediction on some prior information, in this case the target-speaker embedding vector $\bm{e}^{\mathrm{t}}$.
In the literature, this approach has been referred to as \textit{speaker embedding conditioning} \cite{ding20_PVAD1}.
While the VAD and SV tasks are inherently different, they may share some processing steps.
Using a joint model, can be seen as a way of removing redundancy, by letting the model reuse mutual computations.
Therefore, $f_\mathrm{TSVAD}$ can potentially be made more compact than $f_\mathrm{VAD}$ and $f_\mathrm{SV}$ separately, thus making the joint modelling approach of interest for resource constrained applications.
In \cite{ding20_PVAD1}, this approach was shown to yield similar performance to parallel processing, using substantially fewer parameters.

\section{Speaker Conditioning Methods}\label{sec:speaker_conditioning}
In many modelling applications, it is desirable that the prediction of the model can be conditioned on some prior information about the desired output.
For TS-VAD, it is desired to condition the prediction of target speaker voice activity on prior target speaker information, using a target speaker embedding.
This can be done by combining the speech features with the speaker embedding through information fusion.
Specifically, we wish to obtain a vector $\bm{y}_n^\prime$ which incorporates both the information of the input frame $\bm{y}_n \in \mathbb{R}^{D_\mathrm{feat} \times 1}$ and the target speaker information from $\bm{e}^{\mathrm{t}} \in \mathbb{R}^{ D_\mathrm{emb} \times 1}$.
To do this, the information found in the target speaker embedding vector $\bm{e}^{\mathrm{t}}$ must somehow be combined with the input signal $\bm{y}_n$.

\subsection{Concatenation}
To combine the information of speech features and the target speaker embedding, one can simply concatenate them and generate a new space incorporating all information from both vectors. 
Here $\bm{e}^{\mathrm{t}}$ and $\bm{y}_n$ are concatenated such that
\begin{equation}
    \bm{y}_n^\prime = \begin{bmatrix} 
    \bm{y}_n \\
    \bm{e}^{\mathrm{t}}
    \end{bmatrix},
\end{equation}
where $y_n^\prime \in \mathbb{R}^{(D_\mathrm{feat} + D_\mathrm{emb}) \times 1}$ is the concatenated vector of $\bm{e}^{\mathrm{t}}$ and $\bm{y}_n$.

The concatenation approach is a popular information fusion approach, used for many applications, and has also been used for TS-VAD \cite{ding20_PVAD1}. 
A disadvantage is that it expands the dimensionality of the input, leading to a less flexible setup. 
To solve this problem, the concatenated vector can be linearly projected into the original dimension of $\bm{y}$,
\begin{equation}
    \bm{y}_n^\prime = \bm{W} \begin{bmatrix} 
    \bm{y}_n \\
    \bm{e}^{\mathrm{t}}
    \end{bmatrix}  + \bm{b},
\end{equation}
where $\bm{W} \in \mathbb{R}^{D_\mathrm{feat} \times (D_\mathrm{feat} + D_\mathrm{emb})}$ is a projection matrix, $\bm{b} \in \mathbb{R}^{D_\mathrm{feat} \times 1}$ is a bias vector and $y_n^\prime \in \mathbb{R}^{D_\mathrm{feat} \times 1}$ is the concatenated vector projected to a $D_\mathrm{feat}$-dimensional space.

\subsection{Addition}
Another approach is to add the information via addition, which can be interpreted as a biasing of the speech features. 
In the case that $\bm{e}^{\mathrm{t}}$ is of different dimensionality than $\bm{y}_n$, $\bm{e}^{\mathrm{t}}$ can be linearly projected to the dimensionality of $\bm{y}_n$ before addition:
\begin{equation}
    \bm{y}_n^\prime = \bm{y}_n + (\bm{W}\bm{e}^{\mathrm{t}} + \bm{b}), 
\end{equation}
where $\bm{W} \in \mathbb{R}^{D_\mathrm{feat} \times D_\mathrm{emb}}$ and $\bm{b} \in \mathbb{R}^{D_\mathrm{feat} \times 1}$.

This method is more efficient than concatenation, as only $\bm{e}^{\mathrm{t}}$ has to be projected, reducing the number of parameters in $\bm{W}$ by $D_\mathrm{feat}^2$.

The addition method assumes that the dimensionality of the space in which the addition takes place is sufficiently high such that the two vectors are approximately orthogonal, in order to prevent the information from overlapping. This is generally true for DNNs.  

The use of addition to introduce new information is a widely used technique, e.g., it is used in Transformer models to introduce positional information to the input tokens \cite{Vaswani_transformer}. 

\subsection{Multiplication}
Instead of adding the vectors, one could also use a scale the input features through multiplication, by generating a $D_\mathrm{feat}$ scaling vector from the embedding and perform an elementwise multiplication with the input features \cite{perez_film}. 
In practice, this can be done by taking the Hadamard product such that,
\begin{equation}
    \bm{y}_n^\prime= \bm{y}_n \odot (\bm{W}\bm{e}^{\mathrm{t}} + \bm{b}),
\end{equation}
where $\bm{W}$ and $\bm{b}$ has the same dimensions as for addition.

This can be interpreted as a scaling of the input based on $\bm{e}^{\mathrm{t}}$ \cite{perez_film}. 
In this approach, elements of $\bm{y}_n$ that are similar to $\bm{e}^{\mathrm{t}}$ are given a greater weight than elements which are not similar.
However, this assumes that the values $\bm{y}_n$ and $\bm{e}^{\mathrm{t}}$ are somehow meaningfully related.

Whether addition or multiplication conditioning leads to better performance depends on the specific modelling task, although, this can be difficult to determine in practice other than through experiments.

\subsection{Featurewise Linear Modulation}
In an effort to generalize addition and multiplication into a single method, an approach denoted Featurewise Linear Modulation (FiLM) which integrates both methods into a single framework was proposed in \cite{perez_film}.
In this approach, a scaling vector and a bias vector are first computed from $\bm{e}^{\mathrm{t}}$ as follows:
\begin{align}
    \bm{\gamma} &= (\bm{W}_{\bm{\gamma}} \bm{e}^{\mathrm{t}} + \bm{b}_{\bm{\gamma}}),\\ 
    \bm{\beta} &= (\bm{W}_{\bm{\beta}} \bm{e}^{\mathrm{t}} + \bm{b}_{\bm{\beta}}),
\end{align}
where the subscript $\bm{\gamma}$ denotes scaling parameters and $\bm{\beta}$ denotes bias parameters.
These parameters are then used to "modulate" the input $\bm{y}_n$ such that,
\begin{equation}
    \bm{y}_n^\prime = \bm{y}_n \odot \bm{\gamma} + \bm{\beta}.
\end{equation}

FiLM was used for TS-VAD in \cite{ding22_PVAD2} and showed improved performance over concatenation.
While, FiLM makes it possible to have a combination of scaling and biasing, it comes at the cost of additional processing requirements for speaker conditioning.

\subsection{Embedding preprocessing}
One issue that might be encountered using the above-mentioned methods, is that the representation space of the embedding vector $\bm{e}^\mathbf{t}$ might not be compatible with the representation space of the input $\bm{y}_n$.
Passing the speaker embedding through a linear projection before combining it with the input features can alleviate this issue.
However, the "alignment" of representation spaces could require a non-linear transformation.

To solve this issue, the use of a non-linear embedding preprocessing step has been proposed \cite{omalley_conditional_conformer}.
Here, $\bm{e}^\mathrm{t}$ is passed through a non-linear transformation before being combined with $\bm{y}_n$:
\begin{equation}
    \hat{\bm{e}}^{\mathrm{t}} = f_\mathrm{pre}(\bm{e}^\mathrm{t}),
\end{equation}
where $\hat{\bm{e}}^{\mathrm{t}}$ is the preprocessed embedding and $f_\mathrm{pre}$ is some non-linear function.

In \cite{omalley_conditional_conformer}, $f_\mathrm{pre}$ is implemented with a linear up-projection, projecting the input to $2 \times D_\mathrm{emb}$ dimensions, followed by a swish non-linear activation \cite{Ramachandran_swish}, after which another linear projection is used to down-project back to the input dimension $D_\mathrm{emb}$.

While preprocessing $\bm{e}^\mathrm{t}$ leads to a model with more parameters, depending on the specific implementation of $f_\mathrm{pre}$, $f_\mathrm{pre}$ only has to be run once.
As a result, embedding preprocessing does not lead to any significant computational overhead, apart from a larger memory footprint, but might improve performance.



\section{Self-supervised Pretraining}\label{sec:ssl}
In recent years, SSL methods have gained popularity, as they enable learning of highly informative audio representations without the need for labelled data \cite{Abdelrahman_ssl_speech_review}.
Instead of using data labels as the supervisory signal, SSL methods train the model on a \textit{pretext} task, utilizing intrinsic properties of the data as the supervisory signal. 
As an example, audio signals contain some information about how audio signals correlate over time, given by the time-relationship of each sample. 
This information could be used as a supervisory signal. 
For instance, one might mask part of an audio signal and train a model to recover the masked part from the remaining unmasked signal. 
To do this, the model must learn to extract information from the unmasked signal relevant to recovering the masked part. 
Although the pretext task might not be a meaningful \textit{downstream} task (application) itself, the model learns a good representation of audio signals in the process of learning to solve this task \cite{baevski_wav2vec2, HuBERT_paper, Chen_WavLMLS}.
As the information present in the representation learned through SSL is likely to overlap with information needed to solve other audio tasks, the learned audio representation can be reused to solve specific downstream tasks without the need to learn an audio representation from scratch.

Unlabelled data is generally cheap to obtain as opposed to labelled data. 
Therefore, SSL methods enable the use of vast amounts of audio data for pretraining.
This is especially useful when labelled data for the downstream task is limited.
In this case, the SSL model can be used as a starting point for training a model to solve the downstream task. 
Specifically, the pretrained SSL model can be fine-tuned for a specific downstream task using the available labelled data. 
Given that the information in the learned audio representation has a sufficient overlap with the information needed to solve the downstream task, the resulting model will likely have better performance than a model trained from scratch in a purely supervised fashion.

SSL methods have also been found to learn representations which are more robust to out-of-distribution data, than purely supervised methods \cite{hendrycks_ssl_robustness}.
This can be explained by the fact that it enables the model to learn from a much larger and more diverse pool of data.
Additionally, the fact that SSL models are optimized to solve a problem which is related to but not equivalent to the downstream task can help to prevent the model from overfitting.
In \cite{hendrycks_ssl_robustness}, the authors found that besides reducing the need for data labelling, SSL models are also more robust to adversarial examples, label corruption, and input corruptions such as signal noise. This was also observed in \cite{bovbjerg_ssl_pvad, mork2024noise}.

Several SSL methods have been proposed for audio, such as Wav2Vec2 \cite{baevski_wav2vec2}, HuBERT \cite{HuBERT_paper} and WavLM \cite{Chen_WavLMLS}. 
These SSL pretraining tasks assume that the full input \textit{context} is available at all times. 
For example, many frameworks randomly mask segments of an input sequence and use information from all unmasked segments to \enquote{fill in} the masked part.
In this case, the model will learn a non-causal representation of the input.
However, for downstream tasks in which it is desired that the model is causal, such as streaming/real-time VAD.
Therefore, causal pretext tasks are more suitable for pretraining of VAD models.

A recent study found that models trained on a causal SSL pretext task yields better performance for causal \textit{downstream} tasks \cite{mahamied_bpc}.
Some popular causal SSL methods include Contrastive Predictive Coding (CPC) \cite{oord_cpc} and Autoregressive Predictive Coding (APC) \cite{APC_paper}. 
The former uses a contrastive prediction loss, using the feature vector at $k$ time-steps ahead as the attractor and features from a different part of the utterance or from another utterance as the distractor.
Here, only information of previous features are used for the prediction, thus the model learns a causal representation of the input.
Similar to CPC, APC also seeks to predict features $k$ time-steps ahead.
Whereas CPC uses a contrastive loss, APC is a generative approach and seeks to directly predict future input features without the need for negative examples.
Instead, a reconstruction loss, such as $\ell_1$ or $\ell_2$ loss, is used.
APC has been shown to outperform CPC on a number of \textit{downstream} tasks \cite{generative_pretraining_for_speech_with_APC}.

In our previous work, we proposed the use of a causal self-supervised pretraining stage to improve the robustness of a TS-VAD model \cite{bovbjerg_ssl_pvad}.
Here, we pretrained an LSTM-based TS-VAD model using APC, and showed that SSL pretraining can boost the robustness of TS-VAD models in noisy conditions.
Specifically, we pretrained a two layer LSTM model, using logMel filterbank features as input, to predict features $k=3$ time-steps ahead.
We also proposed DenoisingAPC (DN-APC) which extends the APC framework to include denoising, similarly to WavLM.
DN-APC proved to outperform standard APC in both clean and noisy conditions.
More formally, the DN-APC learning objective can be stated as
\begin{equation}
    \mathcal{L}_\text{DN-APC} = \sum_{n=0}^{N-k-1}\norm{g(h(\hat{\bm{y}}_n)) - \bm{y}_{n+k}}_1
\end{equation}
where $\mathcal{L}_\mathrm{DN-APC}$ denotes the DN-APC learning objective, $N$ is the number of input frames, $h$ is the DN-APC encoder, $\hat{\bm{y}}_n$ is the input features computed from an augmented waveform, $k$ denotes how many time-steps ahead the model is trained to predict and $g$ is a regression function which predicts the future frame, $\bm{y}_{n+k}$, which is computed from an unaugmented waveform.

\begin{figure}[tb]
    \centering
    \includegraphics[scale=1.]{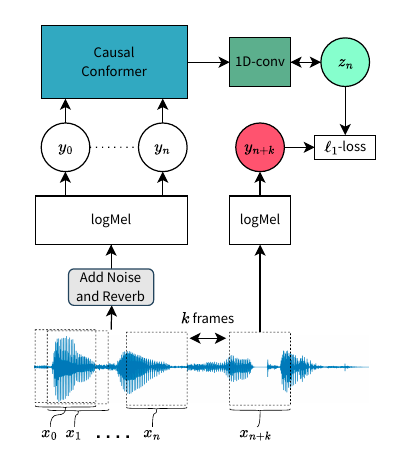}
    \caption{Illustration of the DN-APC framework also used in \cite{bovbjerg_ssl_pvad}.}
    \label{fig:DNAPC}
\end{figure}

For this work, we adopt the same DN-APC framework to pretrain our TS-VAD models.
In our DN-APC implementation, we perform data augmentation on the input by using additive noise and reverberation before computing logMel features.
The prediction label is generated by computing logMel features from the unaugmented clean future speech.
For the regression function, we use a single 1D-convolution layer with a kernel size of 1 and number of channels equal to the feature dimension of $\bm{y}$.
This setup is illustrated in \Cref{fig:DNAPC}.


\section{Experimental Setup}\label{sec:experiments}
We evaluate different model setups and training approaches, all implemented using PyTorch \cite{pytorch}. 
The target-speaker VAD models are evaluated using mean average precision (mAP), commonly used for multi-class classification, as the main evaluation metric. 
Additionally, we present the average precision score for the non-speech (ns), target-speech (ts) and non-target-speech (nts) classes individually.
Noise-robustness is tested by evaluating the models on test sets ranging in SNR from \SIrange{-5}{20}{\decibel} in steps of \SI{5}{\decibel}. Additionally, the models are also tested in clean conditions.

\subsection{Data sets}
As multi-speaker data sets with VAD and speaker identity labels are not readily available, we follow the approach of \cite{ding20_PVAD1} and generate multi-speaker utterances for training and testing.
This is done by selecting an utterance from $n$ speakers, with $n \in \mathcal{U}(1, 3)$, and concatenating them.

In our experiments, we construct both training and test data, using the Librispeech \cite{librispeech} data set, which is a freely available data set consisting of utterances from open-source audiobooks sampled at \SI{16}{\kilo\hertz}.
Librispeech contains approximately \SI{960}{\hour} relatively clean training data with speaker identity information.
This data is split into \textit{clean} and \textit{other} subsets based on how well a reference ASR model performed on the utterances.
Specifically, there are two \textit{clean} sets, one with \SI{100}{\hour} and one with \SI{360}{\hour} of speech.
The remaining \SI{500}{\hour} of speech was named \enquote{other} as it was categorized to be \enquote{less clean}.
A \enquote{clean} and an \enquote{other} set are also available for validation and testing.
When generating multi-speaker utterances, we only use utterances within the same subset, such that utterances from different subsets are not mixed.

\begin{figure*}
    \centering
    \includegraphics[width=\linewidth]{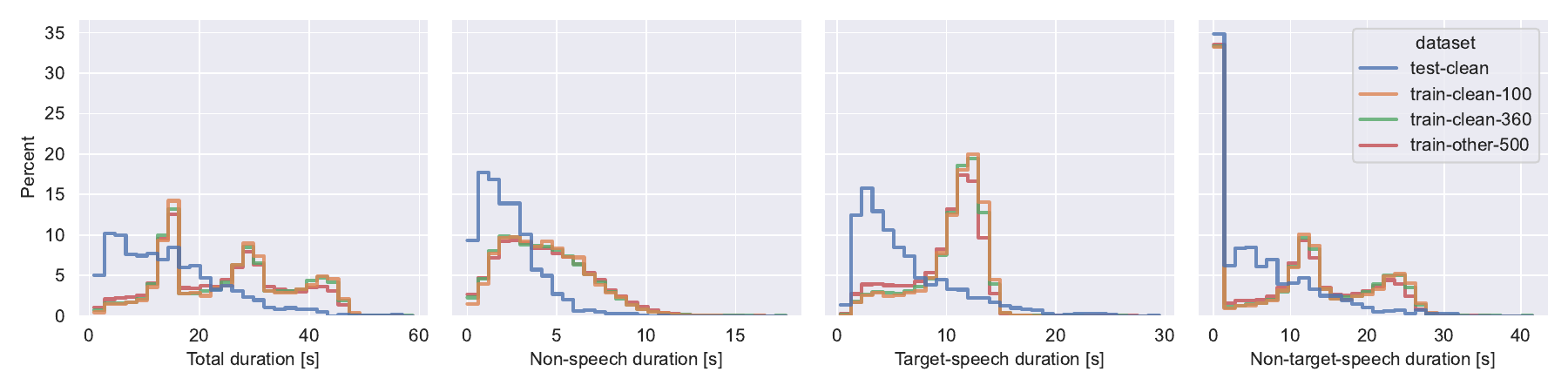}
    \caption{Histograms showing the distributions of the concatenated LibriSpeech training and test set utterances.}
    \label{fig:data_summary}
\end{figure*}

We simulate a setting with a large pool of unannotated data available for pretraining and a smaller pool of labelled data available for supervised training. 
This is done by using train-clean-100, train-clean-360 and train-other-500 during pretraining, while we only use train-clean-100 for supervised training.
The Librispeech metadata can be used to extract a speaker ID, but the data set does not contain VAD labels. 
However, since the speech is relatively clean and has transcriptions, VAD labels can be generated using Forced Alignment, which we do using the Montreal Forced Aligner \cite{montreal_forced_aligner}.
In \Cref{tab:data_summary} an overview of the different data sets used in our experiments is found.
The distributions of utterance durations in the constructed multi-speaker data sets can be seen in \Cref{fig:data_summary}. Here, the distribution of target speech, non-target and non-speech is also shown.

\begin{table}[b]
\centering
\caption{Summary of data sets with concatenated utterances used for training, pretraining and testing. Here, LS denotes LibriSpeech and data set refers to the corresponding subset of the original LibriSpeech data set.}
\label{tab:data_summary}
\footnotesize
\begin{tabular}{@{}llll@{}}
\toprule
Data set          & \#Multi-speaker Utt.    & \#Speakers & Used for \\ \midrule
LS train-100-clean & 9998              & 251  & SSL + Supervised \\
LS train-360-clean & 51943              & 921  & SSL \\
LS train-500-other & 74384              & 1166 & SSL \\
LS test-clean      & 1330               & 40   & Testing \\ \bottomrule
\end{tabular}
\end{table}

For added robustness, we simulate more realistic settings by augmenting the training data with additive noise and reverberation.
We use the noise data set from \cite{Kolbaek_noise_robust_speaker_verification} which contains recordings with babble, bus, café, pedestrian, street and speech-shaped noise.
For reverberation, we use the SLR28 room-impulse-response (RIR) data set also used in \cite{Ko_RIR_data_augmentation}. 
To add noise and reverberation, we use the audiomentations \cite{audiomentations} python package. 

For training, noise is added such that the SNR is between \SIrange{-5}{20}{\decibel}, naturally occurring in speech. 
Noise and reverberation are both added during training, in an online manner, with a probability of \SI{50}{\percent} each.
During training, we keep the café noise as a holdout set for testing in unseen conditions.

When testing the trained models, we only add background noise to the test data.
Whereas the SNR level during training can take on any value between \SIrange{-5}{20}{\decibel}, we fix the SNR level to a specific value in this range during testing. 
Specifically, we test the model performance at SNR levels in this range in steps of \SI{5}{\decibel}. This is done individually for each noise type.

\subsection{TS-VAD model}
Our TS-VAD model takes features from a logMel filterbank feature extractor as input.
The logMel filterbank has 40 Mel-scale filters and has a frame length of $L=400$ and a frame shift of $M=160$, corresponding to \SI{25}{\milli\second} and \SI{10}{\milli\second} of audio, respectively. 
After framing, a Hann window is applied before computing logMel features. 

For audio modelling, we use a causal DNN, consisting of a causal DNN encoder.
First, the logMel features are mapped from the input dimension, $D_\mathrm{input}=40$, to the dimension of the DNN encoder, $D_\mathrm{hidden}$ through a linear layer.
The DNN then encodes the $D_\mathrm{hidden}$-dimensional features into a hidden representation also of dimension $D_\mathrm{hidden}$.
Finally, a classification layer maps the DNN output to three outputs, corresponding to prediction for each output class, i.e., \textit{non-speech} (ns), \textit{target-speech} (ts) and \textit{non-target-speech} (nts).

In our previous work, we used an LSTM model, similar to the one described in \cite{ding20_PVAD1}, as the DNN encoder.
However, self-attention-based models have recently shown very strong performance for many audio modelling tasks.
Inspired by \cite{ding22_PVAD2}, we switch out the LSTM encoder of our previous work for a Conformer encoder with relative positional encoding \cite{dai_transformerXL}. 
Different from \cite{ding22_PVAD2}, our model is smaller. 
Specifically, we use two conformer encoder layers with $D_\mathrm{hidden}=64$ and a single attention head each.
To ensure a causal model, we follow \cite{ding22_PVAD2} and use a causal convolution with a kernel size of 31, and restrict the Conformer context span to 31 frames in the past.
This leads to a context window of \SI{310}{\milli\second} of speech for each prediction.

As the speaker embedding model, we use an off-the-shelve d-vector model \cite{Wan_GE2E} trained on VoxCeleb \cite{NAGRANI_voxceleb}, and LibriSpeech \cite{librispeech} training data.
The d-vector model is a 3-layer LSTM model with a hidden dimension of 256.
It takes 40-dimensional logMel features as input and summarizes the input into a 256-dimensional speaker embedding vector $\bm{e}^{\mathrm{t}}$.
For enrolment of the target speaker, we randomly sample at least \SI{5}{\second} of enrolment speech from each speaker in the data set.
We then generate a target speaker embedding for each speaker using the d-vector model.

\begin{figure*}
    \centering
    \includegraphics[width=\linewidth]{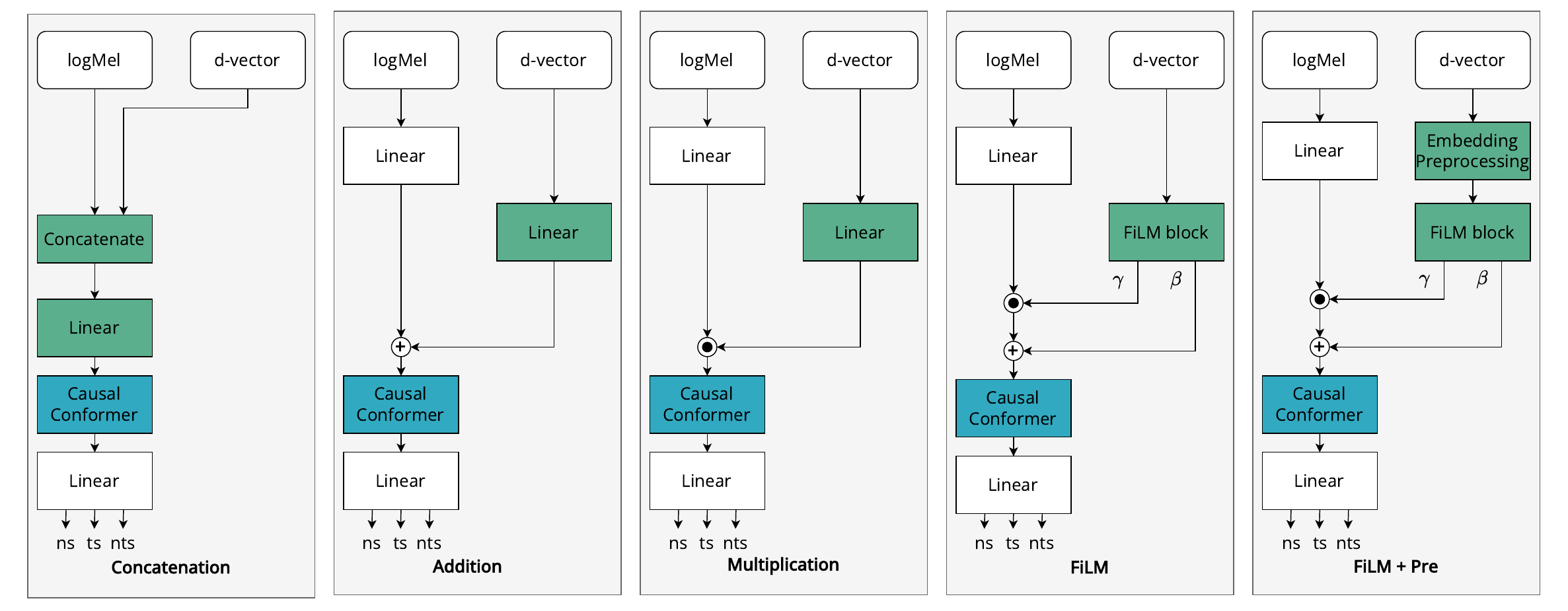}
    \caption{Various TS-VAD model setups for speaker conditioning. Blue boxes designate the pretrained encoder, and green boxes represent elements related to speaker conditioning.}
    \label{fig:model_setups}
\end{figure*}

We carry out experiments using five different model architecture variations using various speaker conditioning methods, to study the effect of speaker conditioning method choice. 
All models are illustrated in \Cref{fig:model_setups}, and a summary of the model sizes is found in \Cref{tab:model_summary}.


\subsubsection{Speaker Embedding Conditioning using concatenation}
Here, we concatenate the input frame features and the speaker embedding vector.
We then project the concatenated vector into the input dimension of the Conformer encoder using a linear layer and encode it using the encoder. 
The output of the Conformer is then passed through a linear classification layer, which outputs the TS-VAD classifications.

\subsubsection{Speaker Embedding Conditioning using addition}
Instead of concatenation, we use addition to condition the model on the target-speaker embedding.
Here, the input frame features and target-speaker embedding are first projected into the input dimension of the Conformer encoder through separate linear layers.
The output of the two linear layers are then added together and encoded using the encoder.
The encoder output is then passed through a linear classification layer, which outputs the TS-VAD classifications.

\subsubsection{Speaker Embedding Conditioning using multiplication}
This model processes the input and speaker embedding similarly as the model using addition. 
However, instead of adding the linearly projected input features and speaker-embedding, their Hadamard product is computed.

\subsubsection{Speaker Embedding Conditioning using FiLM}
This model uses a FiLM block to combine speaker embedding and input frame features.
Here, the input feature vector is first passed through a linear layer followed by a SiLU activation. 
The linear layer up-projects the feature vector into a space with the same dimensionality as the speaker embedding vector. 
The speaker embedding is then passed through a FiLM block to generate $\bm{\gamma}$ and $\bm{\beta}$. 
The factors $\bm{\gamma}$ and $\bm{\beta}$ are then used to scale and bias the up-projected feature vector.
Finally, a linear layer is used to project the resulting vector down into the input dimension of the Conformer encoder.
This vector is then encoded using the encoder and the output is passed through a linear classification layer for TS-VAD classifications.

\subsubsection{Speaker Embedding Conditioning using FiLM w. speaker embedding preprocessing} 
Here, the speaker embedding is first passed through a non-linear preprocessing block before being combined with the input features.
The non-linear block first has a linear layer which projects the embedding into $D_\mathrm{pre} = 2 D_\mathrm{emb}$. 
The up-projected embedding is then passed through a SiLU activation before being projected down into the original dimension.


\begin{table}[tb]
    \centering
    \caption{Summary of TS-VAD model setups.}
    \begin{tabular}{c|cccccc}
    \toprule
         Model & ET1 & ET2 & ET3 & ET4 & ET5\\
         Parameters & 124k & 124k & 124k & 149k & 149k (+263k) \\
    \bottomrule
    \end{tabular}
    \label{tab:model_summary}
\end{table}

All models are trained using cross-entropy loss and AdamW optimizer with a learning rate of 0.001.
We use a cosine annealing with warmup and restart learning rate scheduler \cite{loshchilov_adamw_warmup_restart}.
We train the models for 10 epochs, which was observed to be enough for model convergence, and we save the best model based on the validation set performance.
The scheduler warmup is set to \num{1000} steps, and each scheduler cycle before restart is \num{5000} steps. 
After each cycle, the maximum learning rate is halved and the cycle length is multiplied by 0.9. 
For batching, we use dynamic bucket batching \cite{gonzalez_variable_size_input_batching} with a maximum bucket size of \num{60000} frames, corresponding to a batch size of \SI{1}{\min} of speech.
For computational efficiency, we segment the input utterances into non-overlapping segments with a maximum length of \SI{10}{\second} during training.

\subsection{Self-supervised Pretraining Setup}
In our self-supervised pretraining setup, we reuse the logMel feature extractor and Conformer encoder, from the TS-VAD model.
First, we compute logMel features, $\bm{y}_n$, from the clean input.
Noise and reverberation is then added to the clean input and noisy logMel features, $\hat{\bm{y}}_n$, are computed.
The Conformer encoder then encodes the noisy features into a hidden representation, $\hat{\bm{h}}_n$, and the clean features are used as the prediction labels. 
The encoded noisy input $\hat{\bm{h}}_n$ is passed through a 1D-convolutional layer to generate $\bm{z}_{n}$, which is the predicted clean features, at $k=3$ frames ahead. 
I.e., the clean features $\bm{y}_{n+3}$ are predicted from the noisy features $\hat{\bm{h}}_n$.

We use $\ell_1$-loss between $\bm{z}_{n}$ and $\bm{y}_{n+3}$ as the learning criterion, and otherwise use the same training settings as for supervised training.
However, as more training data is used, we extend the schedule cycle length to \num{30000} steps.
Additionally, we clip the gradients, by nornalizing them to have a maximum 2-norm of 1.

After training, the Conformer encoder weights are copied to the TS-VAD model, which is further fine-tuned.
During fine-tuning, we update all model weights, as freezing the pretrained weights did not yield good performance.

\section{Experimental Results}\label{sec:results}
In the following, we present the results from the described experiments. For comparison of scores, we use the absolute difference in percentage.

\subsection{Effect of DenoisingAPC pretraining}
To evaluate the general effect of pretraining, we compute the performance difference between supervised baseline and pretrained models and take the average over all conditioning methods.
The resulting scores are presented in \Cref{tab:average_improvement}. 
We find that the pretrained model on average shows improvements of \SI{2.15}{\percent} in clean conditions, \SI{2.18}{\percent} in seen conditions and \SI{2.21}{\percent} in unseen conditions.

The performance improvement is slightly higher in low SNR conditions, with the improvement in \SI{-5}{\decibel} being around \SI{0.5}{\percent} higher than in clean conditions. 
This, suggests that pretraining is especially useful for making the model better at distinguishing noise from speech. 

Looking at the average improvement for $\mathrm{ns}$, $\mathrm{ts}$ and $\mathrm{nts}$ individually, we see that all categories are improved.
Here we see, that in clean conditions, the $\mathrm{ts}$ category improves the most with an improvement of \SI{2.86}{\percent}, while the $\mathrm{ns}$ and $\mathrm{nts}$ only improve \SI{1.64}{\percent} and \SI{1.96}{\percent}, respectively.

Interestingly, while the $\mathrm{ns}$ are not improved as much as the $\mathrm{ts}$ and $\mathrm{nts}$ categories for high SNR values, the opposite is true for low SNR.
This can be explained by the fact that it is easier to distinguish noise from speech in high SNR settings than low SNR settings. 
Therefore, the advantage of pretraining in terms of noise robustness is not as prominent for such settings, whereas more challenging settings highlight the improved noise-robustness of the pretrained models.

\begin{table}[h]
\caption{Improvement for pretrained models averaged over all conditioning methods.}
\label{tab:average_improvement}
\centering
\begin{tabular}{@{}llllll@{}}
\toprule
                         &        & \multicolumn{3}{c}{$\Delta$AP [\si{\percent}]} &      \\ \cmidrule{3-5}
SNR                      & Noise  & ns     & ts    & nts   & $\Delta$mAP [\si{\percent}]  \\
\midrule
-5      & Seen   & 3.06   & 2.23  & 2.60  & 2.63 \\
                         & Unseen & 2.64   & 2.51  & 2.30  & 2.48 \\
0       & Seen   & 2.46   & 2.30  & 2.54  & 2.43 \\
                         & Unseen & 2.33   & 2.68  & 2.35  & 2.56 \\
5       & Seen   & 1.89   & 2.29  & 2.12  & 2.10  \\
                         & Unseen & 1.82   & 2.59  & 2.37  & 2.26 \\
10      & Seen   & 1.66   & 2.35  & 1.87  & 1.96 \\
                         & Unseen & 1.55   & 2.52  & 2.04  & 2.04 \\
15      & Seen   & 1.63   & 2.44  & 1.78  & 1.96 \\
                         & Unseen & 1.47   & 2.53  & 1.87  & 1.96 \\
20      & Seen   & 1.65   & 2.52  & 1.77  & 1.98 \\
                         & Unseen & 1.50   & 2.57  & 1.82  & 1.96 \\
Average & Seen   & 2.06   & 2.26  & 2.11  & 2.18 \\
                         & Unseen & 1.89   & 2.57  & 2.18  & 2.21 \\
\midrule
Clean                    & None   & 1.64   & 2.86  & 1.96  & 2.15 \\ \bottomrule
\end{tabular}
\end{table}

\subsection{Effect of speaker conditioning method}
In \Cref{tab:seen_noise} the scores and standard deviation for each individual model, averaged over seen noise types, are given.
The top scoring model for each category is marked in bold.
We see that while all models show improvement from pretraining, the model using FiLM generally performs the best.
However, when looking only at the $\mathrm{ts}$ category, multiplication conditioning performs better.
Additionally, the model using FiLM and embedding preprocessing slightly outperforms using only FiLM for the $\mathrm{ns}$ category. 
From the scores in unseen noise, presented in \Cref{tab:unseen_noise}, a similar picture is seen, although FiLM with speaker embedding preprocessing slightly outperforms only FiLM.

Although FiLM performs best in general, the choice of speaker conditioning method does not seem to have a huge impact on performance in our experiments. 
This suggests that simple conditioning methods such as addition or multiplication are sufficient to condition the encoder input, with multiplication even outperforming more complex methods for the target-speech category. 


\begin{table}[tb]
\caption{Average precision scores for clean speech and speech in seen noise averaged over all SNR levels. The best score for each category is marked in bold. Standard deviation is shown in parentheses.}
\label{tab:seen_noise}
\tiny
\centering
\begin{tabular}{@{}lllllll@{}}
\toprule
           &        &       & \multicolumn{3}{c}{AP [\si{\percent}]}                   &              \\ 
\cmidrule{4-6}
Cond.      & Model       & SNR   & ns          & ts          & nts          & mAP [\si{\percent}]          \\
\midrule
Concat      &  Base      & average & 72.87 (0.42) & 79.11 (0.67) & 80.32 (0.46) & 77.43 (0.52) \\
          \cmidrule{3-7}
           &        & clean & 91.49 (0.21) & 85.94 (0.52) & 88.45 (0.33) & 88.63 (0.35) \\
           \cmidrule{3-7}
        &  DN-APC      & average & 74.30 (0.38) & 79.88 (1.47) & 81.24 (1.01) & 78.47 (0.95) \\
           \cmidrule{3-7}
           &        & clean & 92.41 (0.34) & 87.47 (1.05) & 89.57 (0.67) & 89.81 (0.69) \\
           \midrule
Add        & Base       & average & 74.09 (0.28) & 76.18 (0.87) & 78.42 (0.70) & 76.23 (0.62) \\
           \cmidrule{3-7}
           &        & clean & 92.03 (0.19) & 83.38 (0.70) & 87.00 (0.51) & 87.47 (0.46) \\
           \cmidrule{3-7}
           &  DN-APC  & average & 75.10 (0.33) & 80.67 (0.70) & 82.11 (0.68) & 79.29 (0.57) \\
           \cmidrule{3-7}
           &        & clean & 92.69 (0.20) & 87.50 (0.68) & 89.57 (0.65) & 89.92 (0.51) \\
           \midrule
Mult.      & Base    & average & 70.97 (0.60) & 81.02 (1.26) & 81.12 (0.73) & 77.70 (0.86) \\
           \cmidrule{3-7}
           &        & clean & 88.76 (1.45) & 87.57 (1.38) & 89.41 (0.95) & 88.58 (1.26) \\
           \cmidrule{3-7}
           & DN-APC & average & 73.34 (0.48) & \textbf{81.68} (0.56) & 81.98 (0.62) & 79.00 (0.55) \\
           \cmidrule{3-7}
           &        & clean & 91.49 (0.39) & \textbf{88.01} (0.21) & 89.87 (0.43) & 89.79 (0.34) \\
           \midrule
FiLM.      & Base  & average & 73.87 (0.40) & 79.34 (0.79) & 81.21 (0.57) & 78.14 (0.59) \\
           \cmidrule{3-7}
           &      & clean & 92.15 (0.19) & 85.92 (0.64) & 89.27 (0.42) & 89.11 (0.42) \\
           \cmidrule{3-7}
           & DN-APC  & average & 75.23 (0.64) & 81.07 (0.88) & \textbf{82.32} (0.97) & \textbf{79.54} (0.83) \\
           \cmidrule{3-7}
           &      & clean & \textbf{93.00} (0.33) & 87.87 (0.84) & \textbf{90.12} (0.89) & \textbf{90.33} (0.69) \\
           \midrule
FiLM + Pre & Base   & average & 73.77 (0.68) & 79.55 (0.35) & 81.06 (0.28) & 78.13 (0.44) \\
           \cmidrule{3-7}
           &      & clean & 92.17 (0.13) & 86.04 (0.74) & 88.89 (0.40) & 89.03 (0.43) \\
           \cmidrule{3-7}
           & DN-APC  & average & \textbf{75.39} (0.34) & 80.50 (0.83) & 82.23 (0.34) & 79.37 (0.51) \\
           \cmidrule{3-7}
           &        & clean & 92.97 (0.15) & 86.63 (1.11) & 89.40 (0.67) & 89.67 (0.64) \\
           \bottomrule
\end{tabular}
\end{table}

\begin{table}[tb]
\caption{Average precision scores for unseen noise at varying SNR levels. The best score for each category is marked in bold. Standard deviation is shown in parentheses.}
\label{tab:unseen_noise}
\tiny
\centering
\begin{tabular}{@{}lllllll@{}}
\toprule
           &        &       & \multicolumn{3}{c}{AP [\si{\percent}]}                   &              \\ 
\cmidrule{4-6}
Model      &        & SNR   & ns          & ts          & nts          & mAP [\si{\percent}]          \\
\midrule
Concat.    & Base  & average & 71.09 (0.35) & 77.42 (0.72) & 78.62 (0.48) & 75.71 (0.52) \\
           & DN-APC    & average & 72.29 (0.34) & 78.35 (1.23) & 79.49 (0.86) & 76.71 (0.81) \\
           \midrule
Add        & Base & average & 72.18 (0.19) & 74.23 (0.36) & 76.62 (0.20) & 74.34 (0.25) \\
           & DN-APC   & average & 72.90 (0.33) & 79.00 (0.72) & 80.20 (0.59) & 77.37 (0.55) \\
           \midrule
Mult.      & Base  & average & 69.13 (0.58) & 79.43 (1.41) & 79.53 (0.63) & 76.03 (0.87) \\
           & DN-APC  & average & 71.33 (0.57) & \textbf{80.29} (0.45) & 80.39 (0.55) & 77.34 (0.52) \\
           \midrule
FiLM       & Base & average & 71.51 (0.40) & 77.62 (0.56) & 79.22 (0.53) & 76.12 (0.50) \\
           & DN-APC  & average & 72.86 (0.69) & 79.34 (1.01) & 80.47 (0.89) & 77.56 (0.86) \\
           \midrule
FiLM + Pre & Base  & average & 71.73 (0.55) & 77.72 (0.68) & 79.26 (0.36) & 76.24 (0.53) \\
           & DN-APC   & average & \textbf{73.30} (0.42) & 78.94 (0.72) & \textbf{80.57 }(0.19) & \textbf{77.60} (0.44) \\
\bottomrule
\end{tabular}
\end{table}

\subsection{Representation Analysis}
To get further insights into the model's learned representation, we extract hidden representations from the output of the last Conformer layer. This is done for both clean speech and speech in mixed noise at various SNR levels corresponding to the test SNRs.
For each noise level, we sample 1000 points from each class and combine all representations in a single set, which we transform to two-dimensional representations using t-SNE \cite{vandermaaten_tsne}. After embedding the representations in a two-dimensional space 

\begin{figure}[htb]
    \centering
    \includegraphics[width=\linewidth]{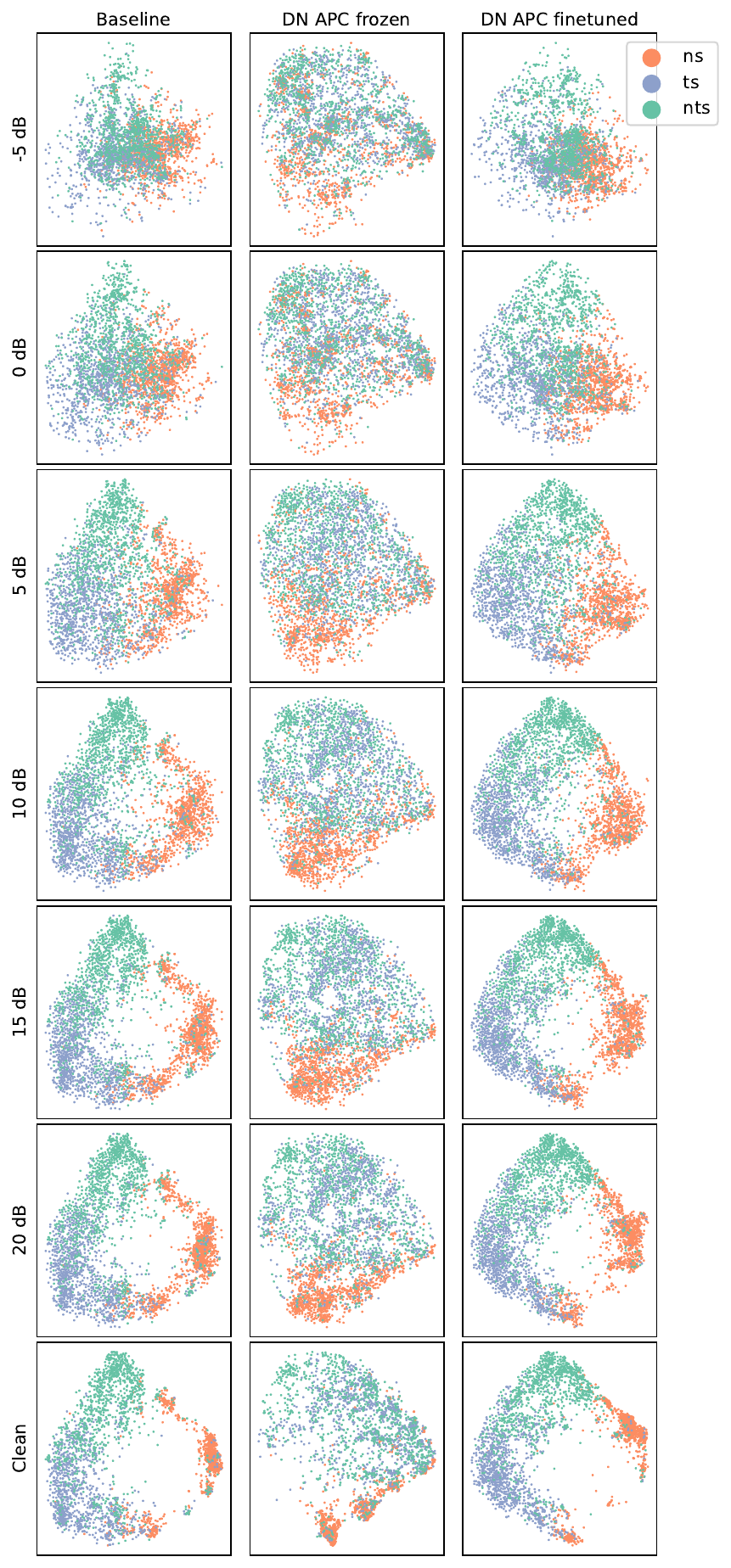}
    \caption{Hidden representation tSNE plot of hidden representations of speech in mixed noise at various dB levels and in clean speech. The first column shows the supervised baseline, the second column shows the DN-APC encoder after pretraining, and the third column shows the DN-APC model after fine-tuning for TS-VAD using FiLM conditioning.}
    \label{fig:representation_in_noise}
\end{figure}

In \Cref{fig:representation_in_noise} tSNE plots of hidden representation for noisy speech at various SNRs of and in clean speech are shown. The colours designate non-speech, target-speech and non-target-speech. 
We show both the supervised baseline, the Conformer encoder after pretraining and the fine-tuned pretrained model.

Generally, we see, that a lower SNR leads to worse separation between classes.
While both the supervised baseline and the fine-tuned pretrained model manage to separate the three classes in clean speech, the fine-tuned model has a slightly better separation in noisy condition.

Looking at the representation from the pretrained model before fine-tuning, we find that the pretrained model has already learned to distinguish noise from speech.
However, the target-speech and non-target speech are randomly mixed.
As the pretrained model does not have any prior information about the target-speaker, this is expected.
However, this suggests that while pretraining cannot provide a good initial representation of target speech vs. non-target speech, it does provide a good representation of speech vs. non-speech. 

\section{Conclusion}\label{sec:conclusion}
In this study, we proposed the use of a causal self-supervised pretraining framework, DenoisingAPC, with the goal of enhancing the performance of speaker-conditioned TS-VAD models and mitigating the need for labelled data.
First, we formally introduced the TS-VAD problem and discussed various model designs as well as speaker conditioning methods.
Five different model setups were proposed, each using a different speaker conditioning method.
We then carried out a series of experiments in which we trained the models and evaluated them in different noisy conditions.
The results showed that pretraining using DenoisingAPC can improve the model performance, especially in noisy conditions, with an average improvement of approximately \SI{2}{\percent} in noisy conditions for both seen and unseen noise.
Furthermore, we found that FiLM conditioning provides the best overall performance, while multiplicative conditioning performs best for detecting target-speech. All proposed conditioning methods showed good performance.
Through a representation analysis, using t-SNE plots of the hidden representations, we determined that while speaker verification is only learned during supervised training, pretraining provides a good initial representation of speech and non-speech, especially for challenging conditions with noise.



\ifCLASSOPTIONcaptionsoff
  \newpage
\fi



\bibliographystyle{IEEEtran}
\bibliography{bibtex/bib/references}

\end{document}